\documentclass{scrartcl}

\usepackage[T1]{fontenc}
\usepackage[utf8]{inputenc}
\usepackage{xcolor}
\usepackage{hyperref}
\usepackage{graphicx}
\usepackage{longtable}
\usepackage{booktabs}
\usepackage{listings}
\usepackage{svg}
\usepackage{authblk}
\usepackage[notes,backend=biber]{biblatex-chicago}
\usepackage[toc,page]{appendix}

\bibliography{references}
\newbox{\myorcidaffilbox}
\sbox{\myorcidaffilbox}{\large\includegraphics[height=1.7ex]{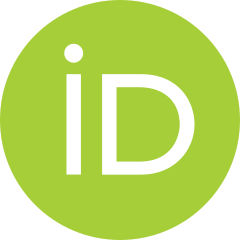}}
\newcommand{\orcidaffil}[1]{%
  \href{https://orcid.org/#1}{\usebox{\myorcidaffilbox}}}

\definecolor{darkblue}{HTML}{004178}
\definecolor{neonblue}{HTML}{009BDC}

\hypersetup{
    colorlinks,
    linkcolor=darkblue,
    citecolor=neonblue,
    urlcolor=neonblue
}

\title{The openCARP CDE}
\subtitle{Concept for and implementation of a sustainable collaborative development environment for research software}

\author[1*]{Felix Bach\orcidaffil{0000-0002-5035-7978}} 
\author[5]{Jochen Klar\orcidaffil{0000-0002-5883-4273}} 
\author[2]{Axel Loewe\orcidaffil{0000-0002-2487-4744}}

\author[2]{Jorge Sánchez\orcidaffil{0000-0002-0824-2691}}

\author[3]{Gunnar Seemann\orcidaffil{0000-0001-7111-7992}}
\author[3]{Yung-Lin Huang\orcidaffil{0000-0003-4307-3907}}
\author[4]{Robert Ulrich\orcidaffil{0000-0001-9063-2703}} 

\affil[1]{Steinbuch Centre for Computing, Karlsruhe Institute of Technology (KIT), Germany}
\affil[2]{Institute of Biomedical Engineering, Karlsruhe Institute of Technology (KIT), Germany}
\affil[3]{Institute for Experimental Cardiovascular Medicine, University Heart Centre Freiburg Bad Krozingen, Germany}
\affil[4]{KIT Library, Karlsruhe Institute of Technology (KIT), Germany}
\affil[5]{jochenklar.de, Germany}
\affil[*]{Correspondence: info@openCARP.org}

\affil[ ]{All authors contributed equally to this work}

\date{\vspace{-1cm}}

\begin{document}

\maketitle

\section*{Abstract}
This work describes the setup of an advanced technical infrastructure for collaborative software development (CDE) in large, distributed projects based on GitLab. We present its customization and extension, additional features and processes like code review, continuous automated testing, DevOps practices, and sustainable life-cycle management including long-term preservation and citable publishing of software releases along with relevant metadata. The environment is currently used for developing the open cardiac simulation software openCARP and an evaluation showcases its capability and utility for collaboration and coordination of sizeable heterogeneous teams. As such, it could be a suitable and sustainable infrastructure solution for a wide range of research software projects.

\section{Introduction}
Successful research software development is often driven by a vibrant user community. Advanced infrastructure can help to make the software maintainable and sustainable\autocite{Anzt2021}. Such an environment was built to foster test-driven development and convenient exchange between users and developers in the DFG-funded SuLMaSS project (Sustainable Life-cycle Management for Scientific Software) and road-tested by the openCARP cardiac electrophysiology simulator\autocite{openCARP,Sanchez-2020-ID14887}. In this work, we describe an infrastructure, which includes a GitLab-based development platform that enables feature requests, bug tracking and release plans, quality assurance practices such as code review by core developers, continuous automated testing, and continuous automated deployment. It also includes tools for sustainable life-cycle management to facilitate reuse and provides means for automatic citable publishing of software releases along with additional data and descriptive metadata in research data repositories, with persistent identifiers (PID) and archiving for long-term availability. Content for the project website is automatically generated and synchronized based on the code repository. All processes are highly automated to reduce manual maintenance effort. The infrastructure could be reused as a development environment for similar research software projects.

\section{Description of Components}

\begin{figure}[p]
\centering
\includegraphics[width=\textwidth]{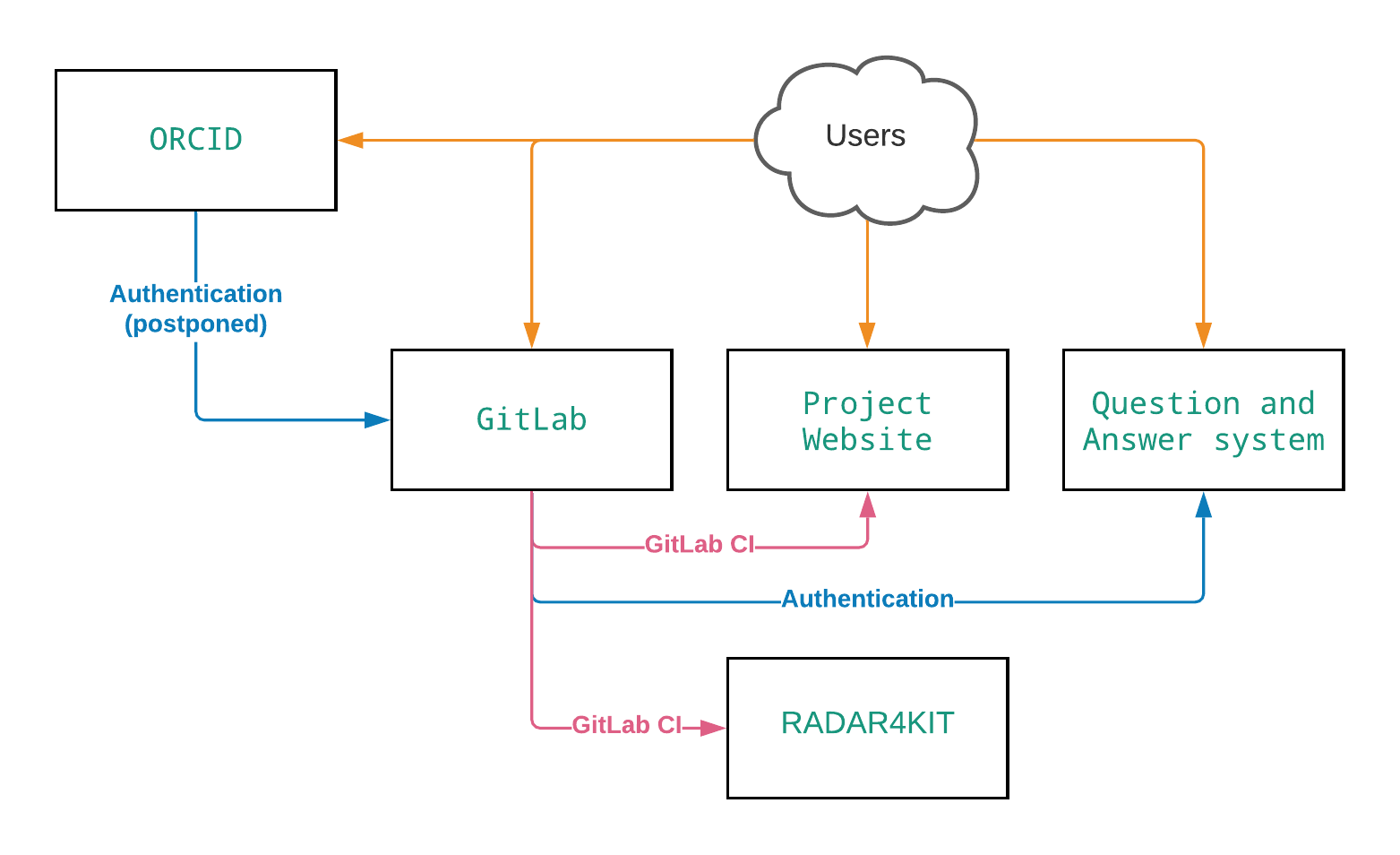}
\caption{Schema of the main components of the openCARP CDE. The users interact with the GitLab software repository, the project web page and the question and answer (Q\&A) system. Authentication for the software repository can be obtained via GitHub/GitLab.com, while the authorization for the Q\&A system is provided by the software repository. RADAR4KIT is used to automatically publish and archive software releases from GitLab}
\label{figure:schema}
\end{figure}

\subsection{Software repository and continuous integration}\label{section:ci}

The core of the openCARP CDE is the software repository\footnote{openCARP GitLab: \url{git.openCARP.org}}, which is based on the open source tool \textbf{GitLab}\footnote{GitLab: \url{https://about.gitlab.com}, accessed July 28, 2021.}. GitLab is a well-established software, which is available both as a hosted service, usable on different free and billed plans, and as self-hosted solution. GitLab offers rich features regarding all stages of the software lifecyle.

GitLab is unique with its feature set as a self-hosted open-source solution for a git-based software development platform. For our purposes, the features of the free Community Edition are sufficient. The software is installed on a virtual machine (VM) running CentOS7 using the GitLab all-in-one package. The specifications of the machine are: 2 CPUs, 8\,GB of RAM and 200\,GB of disk space (mainly used to store the docker images). As an alternative to the self-hosted instance of GitLab, it would be possible to use the as-a-service variant \texttt{GitLab.com} or competing services like \texttt{GitHub} or \texttt{Bitbucket}. While the main functions are similar, a hosted service does not give the same freedom of integration with other services (e.g. regarding authentication, see Section~\ref{section:aa}) and, more important, it implies that all data are hosted on external servers. It also requires a different funding approach, since those services usually bill by user and month.

\begin{figure}[p]
\centering
\includegraphics[width=\textwidth]{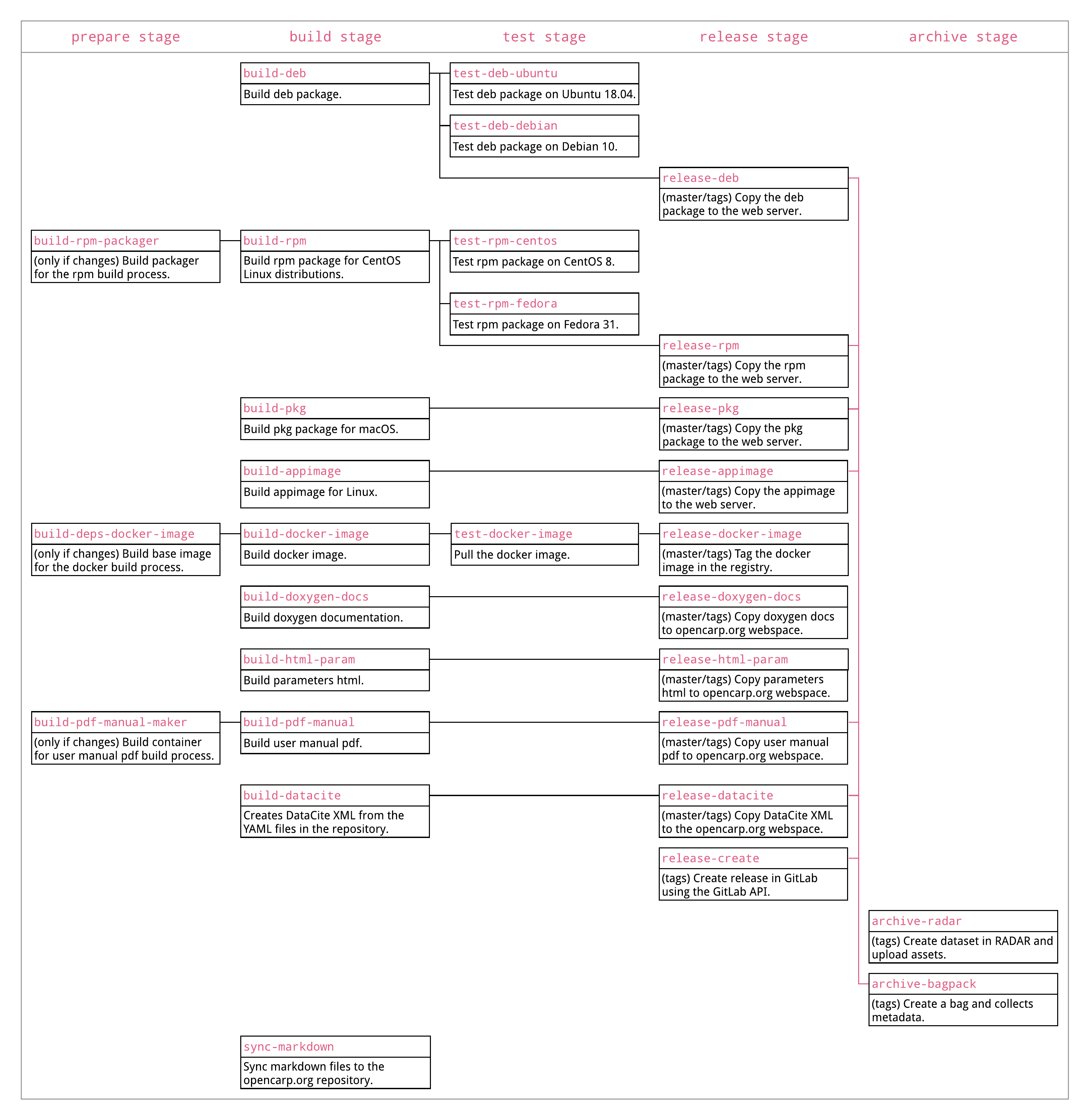}
\caption{Schema of the CI jobs for the openCARP repository. Jobs are separated into different stages (prepare, build, test, release and archive), which are processed sequentially. Dependencies between jobs are indicated by black (same task) or pink (different tasks) arrows. Some jobs are only called for the master branch or only for releases (tagged commits in git).}
\label{figure:ci}
\end{figure}

Integrated within GitLab is a powerful continuous integration (CI) framework, which allows performing automatic tests, generating documentation and other tasks, whenever a new commit is pushed to the repository. Through this functionality we are able to connect the different components of the openCARP CDE and synchronize information between the code repositories, the project webpage and the long-term archive. 

Furthermore, GitLab ships with its own docker registry\footnote{in our case  \url{docker.openCARP.org}}. Docker\footnote{Docker: https://www.docker.com, accessed July 28, 2021.} uses container-based virtualization to encapsulate the whole system environment of an application (except the kernel of the operating system). Using Docker, we can create images for openCARP, which include all prerequisites and can therefore be installed with much less friction than would be the case for distribution of the source code alone.

GitLab offers sophisticated possibilities to create and orchestrate its CI jobs. An overview of our pipelines is given in Figure~\ref{figure:ci}. Jobs can be grouped into different stages, which are processed sequentially, and can have dependencies. The main tasks of the CI for openCARP are:

\begin{itemize}
\item Build \texttt{deb} and \texttt{rpm} packages for Debian, CentOS and Ubuntu based Linux distributions, AppImages for Linux-based systems, as well as \texttt{pkg} packages for macOS. These binary packages allow for an automatic installation of openCARP and all its system prerequisites.
\item Build a Docker image from the openCARP source and upload it to the openCARP docker registry. Docker allows for a frictionless deployment of openCARP on any System which can run Docker (Linux, macOS, Windows).
\item Build the Doxygen\footnote{Doxygen: \url{http://www.doxygen.nl}, accessed July 28, 2021.} documentation of openCARP and copy it to a public directory on openCARP.org\footnote{openCARP documentation: \url{https://openCARP.org/documentation/opencarp-documentation}, accessed July 20, 2021.}. Doxygen automatically creates a browsable documentation from comments (so-called docstrings) in the source code.
\item Build the user manual as pdf and copy it to a public directory on openCARP.org.
\item Build a browsable interface for the different parameters of openCARP and copy it to a public directory on openCARP.org\footnote{openCARP parameters: \url{https://openCARP.org/documentation/opencarp-parameters}, accessed July 28, 2021.}. This and the artifacts mentioned in the two points above are available per git branch and tags and thus allow users to go back to the documentation of any specific version.
\item Build a browseable report interface for the available regression tests and copy the report to the public directory on openCARP.org\footnote{openCARP parameters: \url{https://openCARP.org/documentation/test-reports\#webpage\_testing\_report}, accessed August 04, 2021.}.
\item Generate a DataCite record from structured metadata files in the code repository (see Section~\ref{section:citation}) and copy it to a public directory on openCARP.org.
\item Create releases for certain versions of the code in GitLab. On a technical level, releases are tags in git, which have certain assets assigned to them. In our case, these are the different binary distributions of openCARP. GitLab creates a release page where those assets are accessible, as well as a tarball of the repository (at the release commit).
\item Create a dataset in the long-term archive RADAR4KIT\footnote{RADAR4KIT: \url{https://radar.kit.edu}, accessed July 28, 2021.} and upload the release assets (see Section~\ref{section:citation}).
\item Create a bagpack archive with the release assets and the DataCite metadata file (see Section~\ref{section:citation}).
\item Synchronize the content of the openCARP.org GRAV content management system with structured metadata files from the code repository. Specific pages in the CMS are tagged with \texttt{pipeline} and \texttt{source} metadata entries. For openCARP, if a page has the pipeline \texttt{openCARP}, the CI inserts the content of the file given as source into the metadata of the page. Its content can then be used as structured data by the CMS, e.g. to use the contributors list from the repository to display on the public web page\footnote{Contributors to openCARP: \url{https://openCARP.org/community/contributors}, accessed July 28, 2021.} (see Section~\ref{section:cms} for more information on the GRAV CMS). 
\end{itemize}

Similar, but less extensive pipelines, exist for other repositories, like the \texttt{carputils}\footnote{carputils: \url{https://git.openCARP.org/openCARP/carputils}, accessed July 28, 2021.} framework. Since the compilation process of openCARP causes some computational effort, CI jobs are not executed on the same VM that is used for GitLab but on a different, dedicated VM with 4 CPUs and less disk space. The different assets, which are produced by the CI are copied to a web space under openCARP.org from which they can be downloaded or integrated in the public project web page (see Section~\ref{section:cms}). Besides, the binary packages are uploaded to GitLab package registry, which preserves records between assets and repository. The project-specific pipelines are available in the openCARP git repository. The generic pipelines for creating releases, DataCite records, BagIt/BagPack archive packages, RADAR releases and synchronizing code repository content with the web page are available in the openCARP-CI repository\footnote{openCARP-CI: \url{https://git.openCARP.org/openCARP/openCARP-CI}} together with documentation on how to use them.

\subsection{Project website}\label{section:cms}

For the general public, the web page \texttt{\url{openCARP.org}} acts as the central contact point and hub for information about the software and as starting point for new users and possible collaborators. The page informs about the structure of the consortium, its governance and code of conduct and features news, announcements and recent tweets\footnote{@openCARP Twitter Account: \url{https://twitter.com/openCARP}, accessed July 28, 2021.}. Users can subscribe to an email newsletter to keep informed about openCARP. Code examples and video tutorials make it easy to get started and perform first simulations. The documentation of openCARP and carputils as well as the user manual are embedded into the page.

The page, which is based on GRAV\footnote{GRAV: \url{https://getgrav.org}, accessed July 28, 2021.}, works two-fold: If the CMS content is changed via the GRAV admin interface, the repository in GitLab is automatically updated and if changes are pushed to the master branch of the GitLab repository, the web page is updated accordingly.

In addition to this synchronization between the web page and its git repository, we use the GitLab CI (described in Section~\ref{section:ci}) to integrate additional structured Markdown files from other software repositories. For example, the \href{https://openCARP.org/about/people}{list of contributors} and \href{https://openCARP.org/about/publications}{publications}, and the \href{https://openCARP.org/community/code-of-conduct}{code of conduct}, which are both part of the main openCARP software repository, are synchronized to the web page in this way. The publication list is automatically built from a BibTeX file. In a similar way, the example experiments reside in their own dedicated repository \href{https://openCARP.org/documentation/examples}{web pages} and are created directly from the python source code files of the experiments via a pandoc\footnote{pandoc: \url{https://pandoc.org}, accessed July 28, 2021.} pipeline.

Additionally, usage statistics for the web page (number of requests, visiting countries, visited page, transfer volume) can be created from the local log files without the need for external services or cookies.

\subsection{Question and answer system}

In software engineering, question and answer (Q\&A) platforms have become a valuable tool for developers to discuss, learn, and interact. In recent years, they mostly replaced mailing lists and classical internet forums. The most important Q\&A site nowadays is certainly Stack Overflow\footnote{\url{https://stackoverflow.com/questions}, accessed July 28, 2021.}. Q\&A systems usually employ the following features:

\begin{enumerate}
\item Users are invited to ask a questions using a convenient web interface.
\item Other users can answer these question or comment on the question (e.g., asking for clarification).
\item Specific answers can be marked as accepted if they provide sufficient information to solve the initial problem.
\item All other users are invited to up- or down-vote an answer's quality, thus collectively ordering the thread. This self-regulation nature of the system allows for an effortless access to the best answers in the future.
\item Questions can be tagged with specific topics to keep the system clearly arranged even after a large number of user interactions and to notify the respective maintainers about new questions.
\end{enumerate}

As with the GitLab repository, we decided to have our own, on-premise question and answer solution. Ideally, the platform will grow to a high-quality knowledge library around the openCARP software. Stack Overflow is a centrally hosted platform and set high bars to create new topic-specific instances. Therefore, it was not suitable for this scenario. Several other solutions exist, however:

\begin{itemize}
\item \textbf{Question2Answer} is an open source Q\&A platform using a PHP/MySQL software stack. According to its web page\footnote{\url{https://www.question2answer.org}, accessed July 28, 2021.}, it is used on over 24,000 sites. Question2Answer is licensed under the GPLv2+ license and can therefore be adopted without problems. Its community on GitHub appears quite active, although the last commit to the code in July 2021 was from November 2020.
\item \textbf{Askbot}\footnote{\url{https://askbot.com}, accessed July 28, 2021.} is another open source Q\&A solution and is used, e.g., by the fedora Linux distribution. It uses python and the Django Framework as underlying technology and is licensed under the GPLv3, which allows for an easy adaptation. Its GitHub community seems more active than that of Question2Answer, with commits in recent days. It is based on the no-longer-maintained OQSA system.
\item \textbf{Shapado} was a Ruby on Rails based Q\&A system, but it is clearly unmaintained since 2012. The code is still available on GitHub\footnote{\url{https://github.com/ricodigo/shapado}, accessed July 28, 2021.}.
\item The enterprise software company Atlassian also offers a Q\&A solution as add-on to its collaboration software \textbf{Confluence}\footnote{\url{https://www.atlassian.com/software/confluence/questions}, accessed July 28, 2021.}, which is closed source software and priced depending on the number of users.
\end{itemize}

After weighing up the functions, requirements and maintenance status of the different applications, we decided to use Question2Answer for the openCARP CDE. With PHP, it uses the same underlying technology as the GRAV CMS and MySQL is a well-established and easy-to-maintain database system. While we are not able to integrate Question2Answer seamlessly into openCARP.org in a sustainable way with reasonable effort, the official SnowFlat theme can be customized to follow the color scheme of openCARP.org. Askbot seems to be even better maintained but the different technology stack and smaller installation base tipped the scales.

\subsection{Authentication and authorization}\label{section:aa}

Regarding authentication and authorization, several requirements need to be met for the openCARP CDE:

\begin{itemize}
\item The system should employ uniform user accounts across all components of the CDE, so that users do not need to maintain separate user accounts.
\item The service should be integrated with an external authentication and authorization infrastructure (AAI), removing the need for separate credentials to be created and maintained by the users.
\item The used AAI should already have a wide usage in the academic community, ensuring a high probability that potential collaborators already have an account there.
\item The used AAI needs to be trustworthy, sustainable and fulfill all legal requirements.
\item The system needs to be capable of being integrated into the GitLab instance.
\end{itemize}

As of 2022, there were two systems worth considering with these requirements in mind: connecting to a Shibboleth based AAI or using the OpenID Connect service of ORCID.

\textbf{Shibboleth} is an internet-based single sign-on solution, which allows users to log in to different services using credentials they obtain from the organization they are affiliated with. These services are called Service Providers (SP). When a user tries to authenticate with the SP, a redirect is initiated to another web service, called Identity Provider (IdP), run by the organization of the user. The IdP then decides if the authentication was successful, usually by querying a local directory service (e.g. LDAP), and communicates the result to the SP. If the request was successful, the SP itself issues a query for additional attributes to the IdP (e.g. full name, email address) depending on the particular web service. Finally, the service provider decides if the authentication request was successful and continues with its own logic.

The benefit of this approach is that the user shares its credentials only with the home organization's IdP and the organization can enforce very fine-grained policies on which data are shared with the SP (which might be run by a completely different entity). There is, however, considerable configuration needed to set up such a system. Especially the connection between the SP and the IdP needs to be cryptographically secured. In order to simplify the process, organizations create federations\footnote{e.g., the AAI of the German research network (\url{https://www.aai.dfn.de_aai}), accessed July 28, 2021.} are created to simplify the process even further. Another limitation is that potential users might not be affiliated with home organizations serving as IdPs, for example non-academic users.

Introduced in 2012, the main objective of ORCID is to provide a unique code to identify scientific authors. In order to obtain this personal identifier, researchers need to create an account with the ORCID web page. For this purpose, ORCID was recently recommended by several institutions \autocite[e.g.][]{vierkant2018}. Based on the ORCID iD, ORCID introduced several tools and integrations. One of these services is the possibility to use ORCID as an OpenID Connect Provider.

OpenID Connect\autocite{openid_connect} is part of the OAuth2 authorization framework, which became widely popular with the advent of social networks. Smaller websites use login-with-Google or login-with-facebook buttons to allow users to authorize with their social network accounts and share information or resources (e.g., pictures) from the network with the website. OpenID Connect offers a standardized authentication layer on top of OAuth2.

On a technical level, OpenID Connect works mostly on the client side and involves fewer negotiations between the web service, called Relying Party (RP) and the OpenID Provider (OP). Usually, the OP issues a client ID and a client secret, which needs be configured in the RP. When authenticating with the RP, the user first gets redirected to the OP, much like with Shibboleth. The user logs into the OP, if not already logged in, and acknowledges any authorization of additional information to the RP. Then, the user gets redirected back to the RP with a code added to the HTTP request. Using this code, the RP then issues a request to the OP (authentication itself with the client ID and secret) to obtain the previously authorized user credentials. Based on this information, the RP decides if the authentication was successful and continues with its logic. While large social networks are the most common OPs, ORCID is certainly the most suitable for the academic environment.

Both approaches have their benefits: Shibboleth authentication is common in the academic realm and in particular at universities. Furthermore, its decentralized approach prevents being dependent on a central actor. There is, however a considerable technical overhead in setting up and maintaining a SP and in negotiating the exchange of metadata with different institutions and/or federations. openCARP has potential users all over the world and therefore a large number of IdPs need to be supported. Moreover, not all institutions (e.g. independent research institutions, commercial parties) are part of a Shibboleth federation and would need to apply for user accounts in academic institutions, which is often a long and bureaucratic process.

In contrast, ORCID is a centralized platform. From the point of view of the operators of the CDE, this considerably simplifies the process of integrating an external authentication provider. Since OpenID Connect is a well-known industry standard, the process is very streamlined and well-documented. For users, it is relatively uncomplicated to obtain an ORCID iD, even if they are not affiliated with a research institution and an ORCID iD offers additional advantages as explained above. While using an external site to store identity information raises questions about data privacy, ORCID offers detailed control about which data are stored and shared with the public or with other services \autocite[see also][]{schallaboeck2017}. To be able to receive user's email addresses during account creation, the hosting institution needs to be a paying ORCID member or part of an ORCID consortium\footnote{ORCID membership: \url{https://info.orcid.org/about-membership/}, accessed July 28, 2021.}.

In conclusion, ORCID is the designated choice for the openCARP CDE, since it combines a low entry threshold for an international academic and non-academic audience with manageable maintenance efforts. The privacy issues, which come with using a central platform, are sufficiently addressed. Regarding levels of assurance about the user identity, ORCID meets our rather low requirements.

Unfortunately, as of 2022, ORCID can only be used as authentication provider with GitLab if it is integrated using a customized GitLab package. This package needs to be built again for each new version of GitLab and the process involves several manual steps. This additional effort constitutes a serious challenge for the sustainable operation and maintenance of the whole environment. While we still think that ORCID authentication would be beneficial to the environment, we postponed the integration to minimize the overhead for a sustainable operation of the CDE. Until this issue is resolved, users need to register using their email and a password or using a GitHub or GitLab.com account.

\subsection{Citation and long term preservation}\label{section:citation}

One of the key features of the openCARP CDE is to enable researchers to properly cite the software in their work and thus attributing the effort of the developers and maintainers. In order to be citable, software releases need to be properly archived in a repository and assigned a persistent identifier (PID), preferably a digital object identifier (DOI)\footnote{While DOI is the most common, other systems exist, are discussed, and exhibit different advantages and disadvantages\autocite{katz2019}}. Connected with the creation of a DOI is the collection of descriptive metadata about the dataset (in this case the software release) in form of the DataCite metadata schema\autocite{datacite2019}. These metadata contain information like the title, creators, subjects and publication dates but also information on who contributed to the curation and archiving process.

This practice, initially coming from conventional publications, was recently adopted for research data and is now used for software as well. In this context, the FAIR principles\autocite{wilkinson2016} for research data serve as guiding principles and can be adopted to software, although, for specific characteristics characteristics of software, they need to be extended\autocite{lamprecht2019,Katz2021}. A number of guidelines around software publication and citation supporting this approach have been published \autocite[see][and references therein]{smith2016, scheliga2019, cff}.

For the openCARP infrastructure, these practices result in a set of additional requirements:

\begin{itemize}
\item the software needs to be long-term archived in a suitable repository,
\item a DOI needs to be registered for every release, and
\item additional metadata, which are not directly connected to the development process need to be collected and maintained.
\end{itemize}

In order to be actually able to run the code in the version it was referred to in a publication, it is advisable to not only preserve the source code the release, but also a number of additional assets:

\begin{itemize}
\item the compiled docker image,
\item the compiled binary RPM distribution package (for CentOS systems),
\item the compiled binary DEB distribution package (for Debian or Ubuntu systems),
\item the compiled binary pkg distribution package (for macOS systems),
\item the user manual as PDF
\item additional components like the carputils python framework at the matching revision
\end{itemize}

This release package, which is called submission information package (SIP) in terms of the OAIS reference model\autocite{oais2012} needs to be saved in a way that it can be reliably stored and transferred. A common specification for this kind of packages is BagIt, which was developed by the Library of Congress and is defined in RFC 8493\autocite{bagit2018}. BagIt is a hierarchical file layout convention, which is simple to implement in common file systems. A so-called \textit{bag} directory contains the data directory with the actual \textit{payload}, as well as \textit{tags}, which are files containing technical and organizational metadata, e.g., file checksums or contact email addresses. BagIt can be considered mature and is widely adopted by archives and libraries \autocite[e.g.][]{fritz2014}. In addition, several software libraries supporting the creation and handling exist\footnote{We use \texttt{bagit-python} maintained by the Library of Congress, \url{https://github.com/libraryofcongress/bagit-python}, accessed July 28, 2021.}. Recently, the Research Data Repository Interoperability Working Group of the Research Data Alliance extended BagIt towards BagPack by allowing for content-specific metadata in a dedicated \textit{metadata} directory inside the bag structure\autocite{bagpack2018}. In principle, different extended metadata standards can be added to a BagPack. However, the before-mentioned DataCite metadata schema, which is the minimum information that needs to be provided to get a DOI, is mandatory for BagPack.

\begin{figure}
    \includegraphics[width=\textwidth]{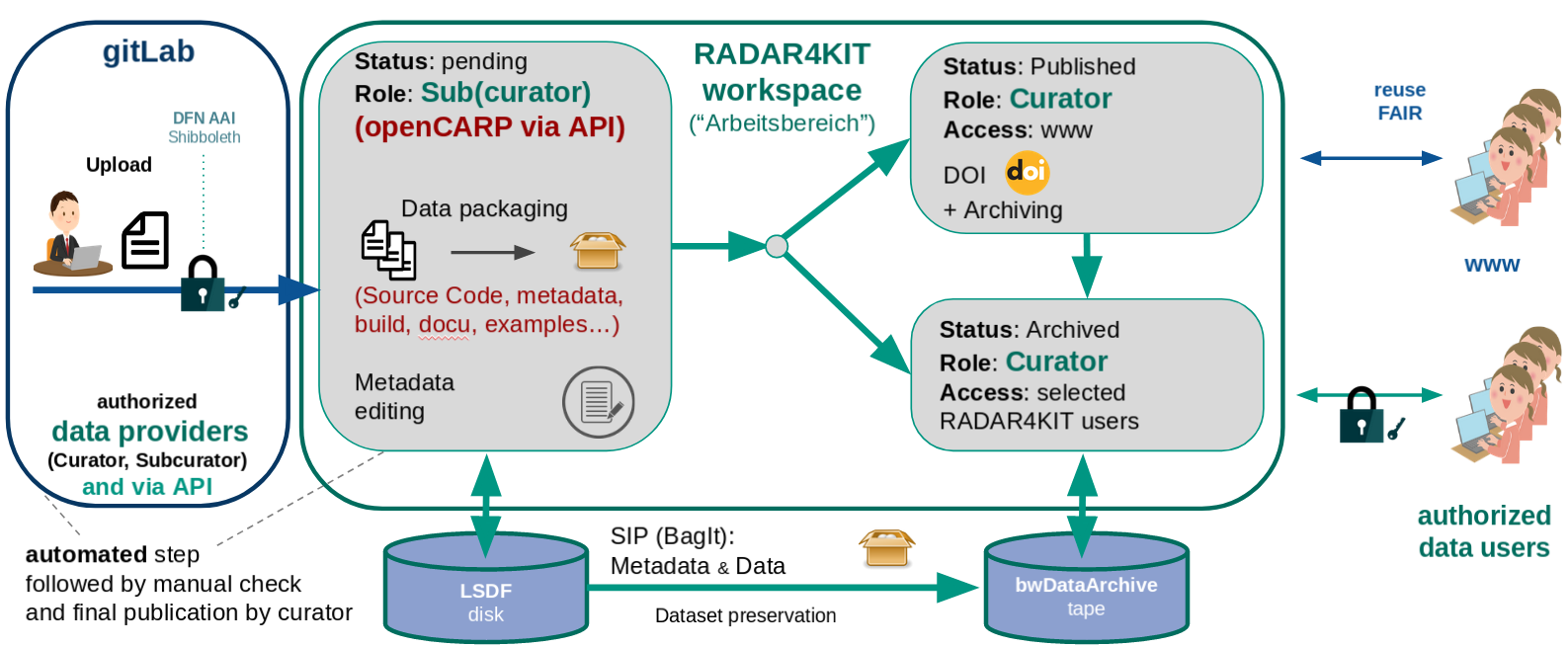}
    \caption{RADAR4KIT publication and archiving workflow of software releases from GitLab repositories.}
    \label{figure:radar4kit}
\end{figure}

For the archival of the created package, several established options were considered. KIT, which hosts the openCARP infrastructure, operates a long-term preservation service called bwDataArchive\footnote{bwDataArchive, \url{https://www.rda.kit.edu/english/}, accessed July 28, 2021.} This service can be used to reliably archive a BagIt or BagPack package. RADAR4KIT\footnote{RADAR4KIT, \url{https://radar.kit.edu}, accessed July 28, 2021.} is operated by KIT as well and is a cooperation with RADAR service\footnote{RADAR, \url{https://www.radar-service.eu}, accessed July 28, 2021.}. It provides a user-friendly web-based workflow to create and archive datasets which is shown in figure \ref{figure:radar4kit}. RADAR uses a metadata schema, which is very similar to DataCite. A possible external service, which is open for all kinds of research data and publications is Zenodo\footnote{Zenodo \url{https://zenodo.org/}, accessed July 28, 2021.}. Zenodo is operated by CERN and was developed under the openAIRE programme. With both RADAR and Zenodo, files are uploaded via HTTP and metadata are added using a web interface or a JSON-based API. In both cases, the creation of the concrete archival package (as stored by the service provider) is not under the control of the user. Since KIT is using RADAR4KIT as a generic institutional repository for research data, we wanted to use it as a depositing solution. This makes it easy to find software contributions by KIT scientists and extract its metadata, e.g. for institutional publication lists.

\begin{itemize}
\item As part of the release process of a particular version of openCARP, the assets described above are automatically uploaded to RADAR4KIT and the corresponding metadata are added using the API of the service. After publication, a DOI is created automatically by the RADAR service and can be used to reference the software release in a text publication or from a different dataset.
\item Optionally or as an alternative, an additional BagPack can be created and archived in the bwDataArchive infrastructure. The resulting package is not publicly available and is used for long-term preservation only.
\end{itemize}

While this strategy relies on infrastructure that is specific to the particular institutional circumstances of the openCARP project, it can be easily adapted to use other infrastructure, e.g., Zenodo instead of RADAR4KIT and any long-term bitstream preservation service instead of bwDataArchive.

In both cases, metadata need to be aggregated as part of the release process. The DataCite schema contains a number of mandatory and optional fields to describe the resource. The fields which we selected for openCARP are given in Table~\ref{table:datacite} of Appendix~\ref{appendix:datacite}. In order to integrate the maintenance of the metadata into the regular software development process, two files were added to the openCARP software repository: the \texttt{CONTRIBUTORS.yml} file contains the list of people involved in the development, publication and preservation of openCARP (Creators and Contributors in the DataCite schema) and the \texttt{METADATA.yml} file contains the other metadata fields. Both files are machine-readable using the YAML format and use naming conventions used in software engineering as well as the DataCite terminology (see Listing~\ref{listing:metadata} of Appendix~\ref{appendix:datacite}). To interoperably identify persons and institutions, we employ the well-established ORCID\footnote{ORCID, \url{https://orcid.org/}, accessed July 28, 2021.} and ROR\footnote{ROR, \url{https://ror.org/}, accessed July 28, 2021.} identification systems.

In order to automatically perform all steps to create the archive package and to aggregate the metadata, we integrate three jobs into the release CI (see also Section~\ref{section:ci}):

\begin{enumerate}
\item Based on the \texttt{CONTRIBUTORS.yml} and the \texttt{METADATA.yml} YAML files in the openCARP repository and the version tag of the release, the \texttt{datacite.xml} file is created and added to the release assets.
\item The release assets are collected and a BagPack archive is created. The archive is copied to in a non-public location in the bwDataArchive infrastructure.
\item Using its REST API, a dataset is is created in RADAR4KIT and metadata is added based on the YAML files in the openCARP repository. In a second step, the assets are uploaded into the dataset. When the curator publishes the dataset with a single click in the RADAR4KIT interface, RADAR4KIT creates a DOI which can be used to cite the release \autocite[e.g.][]{openCARPv5.0}.
\end{enumerate}

\section{Evaluation and Discussion}
The described infrastructure was and is intensively used in the development of the openCARP software, which provides an open cardiac electrophysiology simulation solution\autocite{openCARP}. In the field of cardiac electrophysiology, computational modeling and simulation are becoming increasingly important. openCARP offers single cell as well as multiscale simulations from ion channel to organ level. Additionally, it includes a wide variety of functions for pre- and post-processing of data as well as visualization. The python-based carputils framework enables the user to develop and share simulation pipelines, i.e., automating \textit{in silico} experiments including all modeling/simulation steps. openCARP was derived from the existing closed-source CARPentry solution, which had grown organically over more than 15 years. The software architecture was adapted to meet new requirements and provide a user-friendly interface. The development team comprises contributors and maintainers from several countries. 
As a development platform, the described CDE was introduced in an early stage while still being in development. This allowed early adoptions of user- and developer-requirements and enabled a real-world evaluation of the CDE components and their interplay. The system was intensively used and well accepted by the developer community. The environment proved to serve its purpose with currently more than 250 registered users in the openCARP GitLab instance. 

An increase of resources - such as disk space of the used virtual machine - would ease its operation and make it more scalable for several hundreds of users and multiple simultaneous jobs. A concept for "on-the-fly scaling" could be applied to keep the costs minimal while providing the needed resources. 
Regarding the technical hosting and management of the systems, the costs are relatively low with about 90 Euros per month for the two virtual machines, including a full backup. Maintaining the web server and GitLab requires very little maintenance of around 1 person hour per month for a trained system operator with a good knowledge of the technical systems. We successfully use automatic updates and backups for both the system and GitLab. 
The infrastructure is based on the CentOS Linux 7 distribution, which is supported until mid of 2024. The CDE needs to be adapted or migrated to another distribution then. The used software is based on GitLab, which is open source software, and apart from GitHub, the most popular git repository management software. As expected, it was received well by the developers and users. We use only features available in the free Community Edition of GitLab.

The modular approach of building on existing components and connecting them allows for reuse in other research data management infrastructures. For example, the described concept for citation and long term preservation will be adopted by MO|RE data\footnote{MO|REdata, \url{http://motor-research-data.org}, accessed July 28, 2021.}, a research data repository for sports science.

\section{Conclusion}
The presented CDE framework consisting of a DevOps platform, an interactive project web page and tailored continuous integration pipelines serves the purpose of providing a comprehensive and smooth yet lightweight infrastructure for research software projects. The openCARP CDE combines several existing open pieces of software to an integrated and comprehensive solution. The successful real-world test by the openCARP user and developer community suggests that this collaborative development environment could be a suitable solution for a wide range of research software projects to support sustainability. The modular architecture enables straightforward exchange or addition of components.

\section*{Acknowledgments}
We gratefully acknowledge support by Deutsche Forschungsgemeinschaft (DFG) (project ID 391128822, SuLMaSS) and software contributions of NumeriCor GmbH, Graz. This project has received funding from the European High-Performance Computing Joint Undertaking EuroHPC (JU) under grant agreement No 955495. The JU receives support from the European Union’s Horizon 2020 research and innovation programme and France, Italy, Germany, Austria, Norway, Switzerland governments. We thank Michael Selzer and Philipp Zschumme for developing and providing toolcompendium for generation of user documentation.

\appendix

\section{DataCite metadata for openCARP}\label{appendix:datacite}

In order to (automatically) create DataCite metadata for the openCARP releases, we selected a subset of the DataCite 4.3 kernel to describe the software. The selection is given in Table~\ref{table:datacite}.

While some of the metadata, like the version and the issue data can be automatically extracted from the GitLab system, most of the metadata is provided from a \texttt{METADATA.yml} file which is part of the openCARP git repository. Listing~\ref{listing:metadata} shows the content of this file.

\begin{longtable}[]{llp{8cm}}
\toprule
Id & Property & Description \\
\midrule
\endhead
1      & Identifier & The DOI to be registered. With RADAR, this cannot be determined before creating the archive, it is therefore skipped for the local creation of the BagPack. \\
\midrule
2      & Creator & The list of researchers and developers involved in the creation of openCARP. For each person, properties 2.1-2.5 are collected.\\
2.1    & creatorName & Full name of the creator. \\
2.2    & givenName & Given name of the creator.\\
2.3    & familyName & Family name of the creator.\\
2.4    & nameIdentifier & ORCID iD of the creator.\\
2.4.a  & nameIdentifierScheme & The scheme of identifier for 2.4: \texttt{ORCID}. \\
2.4.b  & schemeURI & The URI of the scheme for 2.4: \url{http://orcid.org}. \\
2.5    & affiliation & The affiliation(s) of the creator.\\
2.5.a  & affiliationIdentifier & The ROR of the affiliation. \\
2.5.b  & affiliationIdentifierScheme & The scheme of identifier for 2.5: \texttt{ROR}. \\
2.5.c  & SchemeURI & The URI of the scheme for 2.5: \url{http://ror.org}. \\
\midrule
3      & Title & The name or title by which the resource is known: openCARP. As the alternative title: Cardiac Electrophysiology Simulator\\
3.a    & titleType & Only used for the alternative title: \texttt{AlternativeTitle} \\
\midrule
4      & Publisher & The institution responsible for the publication of the resource: \texttt{Karlsruhe Institute of Technology (KIT)}.\\
\midrule
5      & PublicationYear & The year of the particular release.\\
\midrule
6      & Subject & The list of subjects for openCARP, taken from the Library of Congress Subject Headings. \\
6.b    & schemeURI & The URL for the Library of Congress Subject Headings: \url{http://id.loc.gov/authorities/subjects}. \\
6.c    & valueURI & The URL for the particular subject, e.g. \url{http://id.loc.gov/authorities/subjects/sh85082124}. \\
\midrule
7      & Contributor & The list of persons and institutions involved in the publication and archival of openCARP. For each person/entity, properties 7.1-7.5 are collected. \\
7.a    & contributorType & Type of contributor from the DataCite controlled vocabulary, e.g. \texttt{DataCurator}, \texttt{HostingInstitution}. \\
7.1    & contributorName & Full name of the contributor.\\
7.2    & givenName & Given name of the contributor.\\
7.3    & familyName & Family name of the contributor.\\
7.4    & nameIdentifier & ORCID iD of the contributor.\\
7.4.a  & nameIdentifierScheme & The scheme of identifier for 7.4: \texttt{ORCID}.\\
7.4.b  & schemeURI & The URI of the scheme for 7.4: \url{http://orcid.org}.\\
7.5    & affiliation & The affiliation(s) of the contributor.\\
7.5.a  & affiliationIdentifier & The ROR of the affiliation.\\
7.5.b  & affiliationIdentifierScheme & The scheme of identifier for 7.5: \texttt{ROR}.\\
7.5.c  & SchemeURI & The URI of the scheme for 7.5: \url{http://ror.org}.\\
\midrule
8      & Date & Important dates for the resource, for openCARP we use \texttt{Created} for the date of the release and \texttt{Issued} for the date of the upload in the archive.\\
8.a    & dateType & The type of date from the DataCite vocabulary: \texttt{Created} or \texttt{Issued}.\\
\midrule
9      & Language & Language of the resource: \texttt{en-US} for American English.\\
\midrule
10     & ResourceType & The type of resource: \texttt{Simulation code}.\\
10.a   & resourceTypeGeneral & The general type of resource from the DataCite vocabulary: \texttt{Software}. \\
\midrule
11     & AlternateIdentifier & The list of altenative identifiers for the resource. We only give the link to the release page in the openCARP GitLab repository: \url{https://git.openCARP.org/openCARP/openCARP/-/releases}.\\
11.a   & alternateIdentifierType & The type of identifier for 11: \texttt{URL}.\\
\midrule
12     & RelatedIdentifier & The list of related identifiers for the resource. We use \texttt{IsNewVersionOf} to link back to an older release of openCARP and \texttt{IsVersionOf} to link to a concept DOI for openCARP.\\
12.a   & relatedIdentifierType & The type of identifier for 12: \texttt{DOI}.\\
12.b   & relationType & The type of relation for 12: \texttt{IsNewVersionOf} and \texttt{IsVersionOf}. \\
\midrule
15     & Version & The version of the openCARP release, identical to the tag in the software repository. \\
\midrule
16     & Rights & The licence for openCARP: "ACADEMIC PUBLIC LICENSE (openCARP, v1.0)" \\
16.a   & rightsURI & The URL where the licence can be found: \url{https://openCARP.org/download/license} \\
\midrule
17     & Description & A short description of openCARP.\\
17.a   & descriptionType & The type of the description for 17: Abstract \\
\midrule
19     & FundingReference & A list containing references about funding. \\
19.1   & funderName & Name of the funder. \\
19.2   & funderIdentifier & ROR of the funder. \\
19.2.a & funderIdentifierType & The type of identifier for 19.2: \texttt{ROR}. \\
19.2.b & SchemeURI & The URI of the scheme for 19.2: \url{https://ror.org}. \\
19.3   & awardNumber & The identifier for the funding grant/award assigned by the funder. \\
19.3.a & awardURI & URL for more information about the funding grant/award. \\
19.4   & awardTitle & Title for the funding grant/award. \\
\bottomrule

\caption{Adoption of the DataCite 4.3 schema for openCARP. DataCite properties not suitable for openCARP are omitted from the table.}
\label{table:datacite}
\end{longtable}

\begin{lstlisting}[caption=METADATA.yml for openCARP, label=listing:metadata]
title: openCARP
additional_titles:
- additional_title: Cardiac Electrophysiology Simulator
  additional_title_type: AlternativeTitle
keywords:
- cardiac electrophysiology
- simulation software
- in silico medicine
- bidomain theory
- computational cardiology
publisher: Karlsruhe Institute of Technology (KIT)
descriptions:
- description: openCARP is an open cardiac electrophysiology simulator for in silico experiments.
  description_type: Abstract
subjects:
- subject: Mathematical models
  value_uri: http://id.loc.gov/authorities/subjects/sh85082124
  scheme_uri: http://id.loc.gov/authorities/subjects
- subject: Computer simulation
  value_uri: http://id.loc.gov/authorities/subjects/sh85029533
  scheme_uri: http://id.loc.gov/authorities/subjects
- subject: Biomedical engineering
  value_uri: http://id.loc.gov/authorities/subjects/sh85014237
  scheme_uri: http://id.loc.gov/authorities/subjects
- subject: Cardiology
  value_uri: http://id.loc.gov/authorities/subjects/sh85020214
  scheme_uri: http://id.loc.gov/authorities/subjects
- subject: Biophysics
  value_uri: http://id.loc.gov/authorities/subjects/sh85014253
  scheme_uri: http://id.loc.gov/authorities/subjects
radar_subjects:
- LifeScience
- ComputerScience
- Medicine
resource: Simulation code
resource_type: Software
alternate_identifiers:
- alternate_identifier: https://openCARP.org/download/releases
  alternate_identifier_type: URL
related_identifiers:
- relation_type: IsVersionOf
  related_identifier:
  related_identifier_type: DOI
- relation_type: IsNewVersionOf
  related_identifier:
  related_identifier_type: DOI
rights: ACADEMIC PUBLIC LICENSE (openCARP, v1.0)
rights_url: https://openCARP.org/download/license
rights_holder: NumeriCor GmbH
funding_references:
- name: Deutsche Forschungsgemeinschaft
  ror: https://ror.org/018mejw64
  award_number: 391128822
  award_uri: https://gepris.dfg.de/gepris/projekt/391128822
  award_title: "Sustainable Lifecycle Management for Scientific Software (SuLMaSS) - Software Dissemination and Infrastructure Development Driven by a Cardiac Electrophysiology Simulator"
\end{lstlisting}

\end{document}